\documentstyle[11pt,epsfig]{article}
\begin{document}
\title{\Large\bf  Canonical interpretation of the $\eta_2(1870)$ }

\author{\small De-Min Li\footnote{E-mail: lidm@zzu.edu.cn}~~and En Wang\\
\small   Department of Physics, Zhengzhou University, Zhengzhou,
Henan 450052, People's Republic of China}
\date{\today}
\maketitle
\vspace{0.5cm}

\begin{abstract}

We argue that the mass, production, total decay width, and decay
pattern of the $\eta_2(1870)$ do not appear to contradict with the
picture of it being the conventional $2\,^1D_2$ $q\bar{q}$ state.
The possibility of the $\eta_2(1870)$ being a mixture of the
conventional $q\bar{q}$ and the hybrid is also discussed.

\end{abstract}

\vspace{0.5cm}


\newpage

\baselineskip 24pt

\section{Introduction}
\indent \vspace*{-1cm}

 In
$\gamma\gamma$ reactions, an isospin zero $2^{-+}$ resonance at
about 1870 MeV in the $\eta\pi\pi$ channel was reported by the
Crystal Ball Collaboration and the CELLO
Collaboration\cite{rr-ex1,cello}. Subsequently, the Crystal Ball
Collaboration presented a more detailed analysis of this resonance
and its mass and width are determined to be about 1881 MeV and 221
MeV, respectively\cite{rr-ex}. In $\bar{p}p$ annihilation, the
Crystal Barrel Collaboration reported an isospin zero $2^{-+}$
resonance with a mass of about 1875 MeV and a width of about 200
MeV in the $f_2(1270)\eta$ channel\cite{ppb}. In radiative
$J/\psi$ decays, the BES Collaboration reported a definite
$2^{-+}$ signal with a mass of about 1840 MeV and a width of about
170 MeV in the $\eta\pi\pi$ channel\cite{jpsi}. In central
production, a similar $2^{-+}$ resonance was observed by the WA102
Collaboration in the $a_2(1320)\pi$, $f_2(1270)\eta$, and
$a_0(980)\pi$ channels\cite{wa1021,wa1022,wa1023}. Using high
statistics data on $\bar{p}p\rightarrow \eta\pi^0\pi^0\pi^0$,
Anisovich et al. confirmed the presence of an isospin zero
$2^{-+}$ resonance with a mass of about 1860 MeV and a width of
about 250 MeV in the $a_2(1320)\pi$, $f_2(1270)\eta$, and
$a_0(980)\pi$ channels\cite{ppb1}. It has been established that
these observations in different experiments refer to a single
state $\eta_2(1870)$\cite{bugg, pdg08}, although this state is
stated to need confirmation\cite{pdg08}. The mass and width of the
$\eta_2(1870)$ are quoted to be  $1842\pm 8$ MeV and $225\pm 14$
MeV, respectively\cite{pdg08}.

With the $\eta_2(1645)$ as the well-established $1\,^1D_2$
$q\bar{q}$ state\cite{pdg08}, the $\eta_2(1870)$ looks like either
a $1\,^1D_2$ $s\bar{s}$ [$\eta_2(1D s\bar{s})$] or a $n\bar{n}$
hybrid [$\eta_2(H n\bar{n})$] ($n\bar{n}\equiv
(u\bar{u}+d\bar{d})/\sqrt{2}$) based on its mass, because the
observed mass of the $\eta_2(1870)$ just overlaps the
Godfrey-Isgur quark model prediction of 1.89 GeV for the
$\eta_2(1D s\bar{s})$\cite{GI} and the flux-tube model prediction
of $1.8-1.9$ GeV for the $\eta_2(H n\bar{n})$\cite{ftmass}. Of
course the strong $a_2(1320)\pi$ and $f_2(1270)\eta$ modes are not
expected from $s\bar{s}$ and hence may imply large
$n\bar{n}\leftrightarrow s\bar{s}$ flavor mixing in the
$\eta_2(1870)$ if the $\eta_2(1870)$ is indeed a $1\,^1D_2$
$q\bar{q}$ state. The near degeneracy of the $\eta_2(1645)$ and
$\pi_2(1670)$ suggests the ideal mixing in the $1\,^1D_2$ meson
nonet, which disfavors the large flavor mixing in the
$\eta_2(1870)$ qualitatively. Also, the calculations from the
$^3P_0$ model quantitatively argue against assigning the
$\eta_2(1870)$ to the $1\,^1D_2$ $n\bar{n}\leftrightarrow
s\bar{s}$ mixed quark model state\cite{strangebarnes}. A feature
of $\bar{p}p$ annihilation is that the well-known $s\bar{s}$ state
such as the $f^\prime_2(1525)$ are produced very weakly, if at
all\cite{bugg}; the $\eta_2(1870)$, in contrast, is produced
strongly, which makes the $\eta_2(1D s\bar{s})$ interpretation for
the $\eta_2(1870)$ unlikely. Therefore, both the decay modes and
production information for the $\eta_2(1870)$ do not favor it
being the $\eta_2(1D s\bar{s})$. Apart from the $\eta_2(1870)$
mass, it dominantly decaying to the $a_2(1320)\pi$ and
$f_2(1270)\eta$ is also in accord with the flux-tube model
expectation for the $\eta_2(H n\bar{n})$ where the preferred decay
channels are to $P+S$-wave pairs\cite{close1,nnhybridbarnes,pss}.
In addition, the discovery of the $\eta_2(2030)$ in $\bar{p}p$
annihilation\cite{ppb1,ppb2} to some extent leaves the
$\eta_2(1870)$ as an `extra', i.e., non-$q\bar{q}$
state\cite{ppb1} since the $\eta_2(2030)$ looks like a natural
candidate for the $\eta_2(1645)$'s first radial excitation from
its mass which is close to the Godfrey-Isgur quark model
prediction of 2.13 GeV\cite{GI} for the $2\,^1D_2$ $n\bar{n}$
state [$\eta_2(2D n\bar{n})$]. So, the hybrid interpretation for
the $\eta_2(1870)$ becomes a popular
opinion\cite{rr-ex,ppb1,bugg,strangebarnes,close1,nnhybridbarnes,pss,amsler,klempt}.
Apart from the $\eta_2(1870)$, its companion $\pi_2(1880)$ was
also regarded to be a viable non-exotic hybrid
candidate\cite{bugg,strangebarnes,close1,nnhybridbarnes,pss,amsler,klempt,anisovich20012,e8521,e8522,e8523}.

 Although the hybrid interpretation
for the $\pi_2(1880)$ and $\eta_2(1870)$ has several attractive
features, it is necessary to exhaust their possible conventional
$q\bar{q}$ descriptions before resorting to more exotic
interpretations. In fact, the observation of the $\pi_2(1880)$ in
the $\rho\omega$ and the $f_2(1270)\pi$ $D$-wave channels strongly
casts doubt over the hybrid interpretation for the $\pi_2(1880)$
since the $\rho\omega$ is expected to vanish and the
$f_2(1270)\pi$ $D$-wave is strongly suppressed for the
hybrid\cite{pss}. In our previous work\cite{lidm2009}, we argued
that the experimental evidence for the $\pi_2(1880)$ is consistent
with it being the conventional $2\,^1D_2$ meson rather than the
$2^{-+}$ light hybrid by investigating its strong decay
properties. If the $\pi_2(1880)$ can be described as the ordinary
$2\,^1D_2$ meson, one natural question is whether its companion
$\eta_2(1870)$ could also be the ordinary $2\,^1D_2$ meson or not.
In this work, we shall discuss the possibility of the
$\eta_2(1870)$ being the $\eta_2(2D n\bar{n})$ from its mass,
production, total width, and strong decay pattern.

The organization of this paper is as follows. In Sections 2-3, we
discuss the mass and production properties of the $\eta_2(1870)$.
In Sect. 4, after a brief review of the $^3P_0$ model and the
flux-tube model used in this work, we present the partial decay
widths of the $\eta_2(1870)$ as the $\eta_2(2D n\bar{n})$ within
these two models. The discussions and conclusion are given in
Sections 5-6, respectively.

\section{Mass}
\indent\vspace{-1cm}

Godfrey-Isgur quark model predicted that the $\eta_2(2D n\bar{n})$
mass is about 2.13 GeV\cite{GI}, about 250 MeV higher than the
$\eta_2(1870)$ mass. The $\eta_2(1870)$ therefore appears too
light to be the $\eta_2(2D n\bar{n})$ at first glance. However, it
should be noted that the $a_1(1700)$ and $a_2(1700)$, both about
100-200 MeV lower in mass than the Godfrey-Isgur quark model
anticipated\cite{GI}, turn out the excellent candidates for radial
excitations\cite{nnhybridbarnes,pss}, which indicates that
Godfrey-Isgur quark model maybe overestimates the masses of the
higher-$L$ radially excited mesons by about 100-200
MeV\cite{GImodel}. So the $\eta_2(2D n\bar{n})$ with a mass about
1.9 GeV is presumably not implausible. Also, the isovector state
should act as a beacon for the mass scale of a meson nonet. If the
$\pi_2(1880)$ can be identified as the isovector member of the
$2\,^1D_2$ $q\bar{q}$ nonet\cite{lidm2009}, the $\eta_2(2D
n\bar{n})$ would be the orthogonal partner of the $\pi_2(1880)$
and one can naturally expect that the $\eta_2(2D n\bar{n})$
degenerates with the $\pi_2(1880)$ in effective quark masses. The
similar behavior also exists in the established $1\,^1D_2$ and
$1\,^3D_3$ meson nonets\cite{pdg08}. Recently, different
approaches already consistently suggested that the $\eta_{2}(2D
n\bar{n})$ has a mass of about 1.9 GeV, close to the
$\eta_2(1870)$ mass. For example, the Vijande-Fernandez-Valcarce
quark model predicted $M_{\eta_2(2Dn\bar{n})}$ = 1.863
GeV\cite{vfv}, the spectrum integral equation expected
$M_{\eta_2(2Dn\bar{n})}$ = 1.937 GeV\cite{spectrum}, and the
Mezoir-Gonzalez quark model found $M_{\eta_2(2Dn\bar{n})}$ = 1.913
GeV\cite{mg}. Therefore, the assignment of the $\eta_2(1870)$ as
the $\eta_2(2D n\bar{n})$ does not appear to be irrational based
on its mass.

\section{ Production}
\indent\vspace{-1cm}

For central production, Close and Kirk have found a kinematic
filter that seems to suppress the well-established $q\bar{q}$
states when they are in $P$ and higher waves\cite{filter}. Its
essence is that the pattern of resonances produced in the central
production process depends on
$dp_T=|\vec{k}_{T_1}-\vec{k}_{T_2}|$, the vector difference of the
transverse momentum recoil of the final state protons. It has been
illustrated in several channels that for $dp_T$ large the
$q\bar{q}$ states are prominent whereas for $dp_T$ small all the
undisputed $q\bar{q}$ states are suppressed while the enigmatic
states probably having more complex structures such as the
$f_0(1500)$, $f_0(1700)$, and $f_0(980)$ survive\cite{ex-filter}.
The application of this kinematic filter to the centrally produced
$K\bar{K}\pi$ system, where the $f_1(1285)$ and $f_1(1420)$ have
the same behavior as a function of the $dp_T$, successfully
established the $f_1(1420)$ as the $^3P_1$ $q\bar{q}$
states\cite{zphysc97}. At one time, the $f_1(1420)$ was
interpreted as either a hybrid\cite{expl1}, a four quark
state\cite{expl2}, or a $K^\ast K$ molecule\cite{expl3}.

In central production both the $\eta_2(1645)$ and the
$\eta_2(1870)$ were clearly observed, furthermore they exhibit the
same behavior as a function of the $dp_T$, appearing sharply when
$dp_T>0.5$ GeV, and vanishing as $dp_T\rightarrow 0$ GeV
\cite{wa1021,wa1022} (see Table 2 of Ref.\cite{wa1021} and Table 2
of Ref.\cite{wa1022}), as do other well-established $q\bar{q}$
states such as the $f_1(1285)$ and $f_1(1420)$. This strongly
suggests that the $\eta_2(1870)$ has the same dynamical structure
as the $\eta_2(1645)$, namely the standard $2^{-+}$ $q\bar{q}$. As
mentioned above, the production in $\bar{p}p$ annihilation process
argues against the $s\bar{s}$ interpretation of the
$\eta_2(1870)$. Therefore, with the $\eta_2(1645)$ as the
well-established $1\,^1D_2$ $n\bar{n}$ state, the production
properties of the $\eta_2(1870)$ are consistent with it being the
$\eta_2(2Dn\bar{n})$.

\section{ Decay}
\subsection{The $^3P_0$ model and the flux-tube model}
\indent\vspace{-1cm}

The $^3P_0$ model and the flux-tube model which are the standard
models for strong decays at least for mesons in the initial state,
have been widely used to evaluate the strong decays of
hadrons\cite{strangebarnes,nnhybridbarnes,lidm2009,3p0rev1,3p0rev2,3p0rev3,3p0rev4,3p0rev5,flux,3p0sho1,3p0sho2,prd72,3p04},
since they give a good description of many of the observed decay
amplitudes and partial widths of the hadrons. Below, we shall give
the brief review of the two models employed in this work.

\subsubsection{The $^3P_0$ model of meson decay} \indent
\vspace*{-1cm}

 The $^3P_0$ model, also known as the quark-pair creation
model, was originally introduced by Micu\cite{micu} and further
developed by Le Yaouanc et al.\cite{3p0rev1}.  The main assumption
of the $^3P_0$ model of meson decay is that strong decays take
place via the creation of a $^3P_0$ quark-antiquark pair from the
vacuum. The newly produced quark-antiquark pair ($q_3\bar{q}_4$),
together with the $q_1\bar{q}_2$ within the initial meson,
regroups into two outgoing mesons in all possible quark
rearrangement ways, which corresponds to the two decay diagrams as
shown in Fig.1 for the meson decay process $A\rightarrow B+C$.

\vspace*{-0.5cm}
\begin{figure}[hbt]
\begin{center}
\epsfig{file=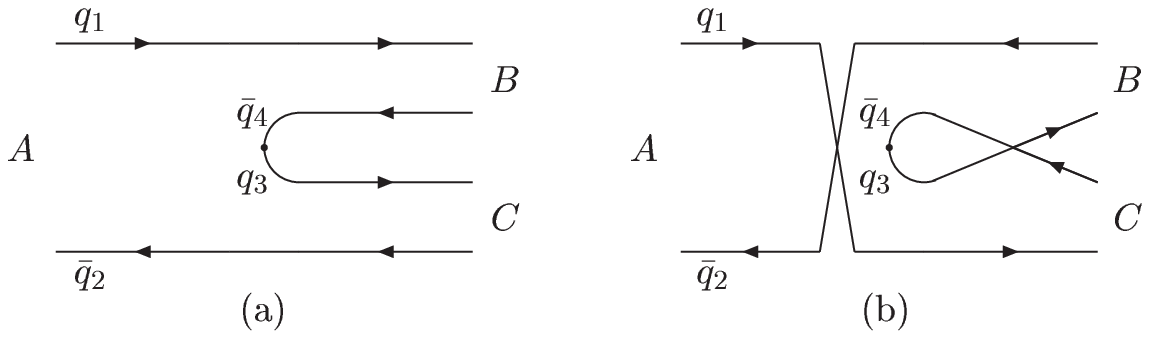,width=6.0cm, clip=}
\vspace*{0.5cm}\vspace*{-0.8cm}
 \caption{\small The two possible diagrams contributing to $A\rightarrow B+C$ in the $^3P_0$
 model.}
\end{center}
\end{figure}

The transition operator $T$ of the decay $A\rightarrow BC$ in the
$^3P_0$ model is given by
\begin{eqnarray}
T=-3\gamma\sum_m\langle 1m1-m|00\rangle\int
d^3\vec{p}_3d^3\vec{p}_4\delta^3(\vec{p}_3+\vec{p}_4){\cal{Y}}^m_1\left(\frac{\vec{p}_3-\vec{p}_4}{2}\right
)\chi^{34}_{1-m}\phi^{34}_0\omega^{34}_0b^\dagger_3(\vec{p}_3)d^\dagger_4(\vec{p}_4),
\end{eqnarray}
where $\gamma$ is a dimensionless parameter representing the
probability of the quark-antiquark pair $q_3\bar{q}_4$ with
$J^{PC}=0^{++}$ creation from the vacuum, and $\vec{p}_3$ and
$\vec{p}_4$ are the momenta of the created quark $q_3$ and
antiquark  $\bar{q}_4$, respectively. $\phi^{34}_{0}$,
$\omega^{34}_0$, and $\chi_{{1,-m}}^{34}$ are the flavor, color,
and spin wave functions of the  $q_3\bar{q}_4$, respectively. The
solid harmonic polynomial
${\cal{Y}}^m_1(\vec{p})\equiv|p|^1Y^m_1(\theta_p,\phi_p)$ reflects
the momentum-space distribution of the $q_3\bar{q}_4$ .

For the meson wave function, we adopt the mock meson
$|A(n^{2S_A+1}_AL_{A}\,\mbox{}_{J_A M_{J_A}})(\vec{P}_A)\rangle$
defined by\cite{mock}
\begin{eqnarray}
|A(n^{2S_A+1}_AL_{A}\,\mbox{}_{J_A M_{J_A}})(\vec{P}_A)\rangle
&\equiv& \sqrt{2E_A}\sum_{M_{L_A},M_{S_A}}\langle L_A M_{L_A} S_A
M_{S_A}|J_A
M_{J_A}\rangle\nonumber\\
&&\times  \int d^3\vec{p}_A\psi_{n_AL_AM_{L_A}}(\vec{p}_A)\chi^{12}_{S_AM_{S_A}}\phi^{12}_A\omega^{12}_A\nonumber\\
&&\times  \left|q_1\left({\scriptstyle
\frac{m_1}{m_1+m_2}}\vec{P}_A+\vec{p}_A\right)\bar{q}_2
\left({\scriptstyle\frac{m_2}{m_1+m_2}}\vec{P}_A-\vec{p}_A\right)\right.\rangle,
\end{eqnarray}
where $m_1$ and $m_2$ are the masses of the quark $q_1$ with a
momentum of $\vec{p}_1$ and the antiquark $\bar{q}_2$ with a
momentum of $\vec{p}_2$, respectively. $n_A$ is the radial quantum
number of the meson $A$ composed of $q_1\bar{q}_2$.
$\vec{S}_A=\vec{s}_{q_1}+\vec{s}_{\bar{q}_2}$,
$\vec{J}_A=\vec{L}_A+\vec{S}_A$, $\vec{s}_{q_1}$
($\vec{s}_{\bar{q}_2}$) is the spin of $q_1$ ($\bar{q}_2$), and
$\vec{L}_A$ is the relative orbital angular momentum between $q_1$
and $\bar{q}_2$. $\vec{P}_A=\vec{p}_1+\vec{p}_2$,
$\vec{p}_A=\frac{m_2\vec{p}_1-m_1\vec{p}_2}{m_1+m_2}$. $\langle
L_A M_{L_A} S_A M_{S_A}|J_A M_{J_A}\rangle$ is a Clebsch-Gordan
coefficient, and $E_A$ is the total energy of the meson $A$.
$\chi^{12}_{S_AM_{S_A}}$, $\phi^{12}_A$, $\omega^{12}_A$, and
$\psi_{n_AL_AM_{L_A}}(\vec{p}_A)$ are the spin, flavor, color, and
space wave functions of the meson $A$, respectively. The mock
meson satisfies the normalization condition
\begin{eqnarray}
\langle A(n^{2S_A+1}_AL_{A}\,\mbox{}_{J_A M_{J_A}})(\vec{P}_A)
|A(n^{2S_A+1}_AL_{A}\,\mbox{}_{J_A
M_{J_A}})(\vec{P}^\prime_A)\rangle=2E_A\delta^3(\vec{P}_A-\vec{P}^\prime_A).
\end{eqnarray}
The $S$-matrix of the process $A\rightarrow BC$ is defined by
\begin{eqnarray}
\langle BC|S|A\rangle=I-2\pi i\delta(E_A-E_B-E_C)\langle
BC|T|A\rangle,
\end{eqnarray}
with
\begin{eqnarray}
\langle
BC|T|A\rangle=\delta^3(\vec{P}_A-\vec{P}_B-\vec{P}_C){\cal{M}}^{M_{J_A}M_{J_B}M_{J_C}}_{(
^3P_0)},
\end{eqnarray}
where ${\cal{M}}^{M_{J_A}M_{J_B}M_{J_C}}_{(^3P_0)}$ is the
helicity amplitude of $A\rightarrow BC$. In the center of mass
frame of meson $A$, ${\cal{M}}^{M_{J_A}M_{J_B}M_{J_C}}_{(^3P_0)}$
can be written as
\begin{eqnarray}
{\cal{M}}^{M_{J_A}M_{J_B}M_{J_C}}_{(^3P_0)}(\vec{P})&=&\gamma\sqrt{8E_AE_BE_C}
\sum_{\renewcommand{\arraystretch}{.5}\begin{array}[t]{l}
\scriptstyle M_{L_A},M_{S_A},\\\scriptstyle M_{L_B},M_{S_B},\\
\scriptstyle M_{L_C},M_{S_C},m
\end{array}}\renewcommand{\arraystretch}{1}\!\!
\langle L_AM_{L_A}S_AM_{S_A}|J_AM_{J_A}\rangle\nonumber\\
&&\times\langle L_BM_{L_B}S_BM_{S_B}|J_BM_{J_B}\rangle\langle
L_CM_{L_C}S_CM_{S_C}|J_CM_{J_C}\rangle\nonumber\\
&&\times\langle 1m1-m|00\rangle\langle
\chi^{14}_{S_BM_{S_B}}\chi^{32}_{S_CM_{S_C}}|\chi^{12}_{S_AM_{S_A}}\chi^{34}_{1-m}\rangle\nonumber\\
&&\times[f_1I_{(^3P_0)}(\vec{P},m_1,m_2,m_3)\nonumber\\
&&+(-1)^{1+S_A+S_B+S_C}f_2I_{(^3P_0)}(-\vec{P},m_2,m_1,m_3)],
\label{3p0exp}
\end{eqnarray}
with $f_1=\langle
\phi^{14}_B\phi^{32}_C|\phi^{12}_A\phi^{34}_0\rangle$ and $f_2=
\langle\phi^{32}_B\phi^{14}_C|\phi^{12}_A\phi^{34}_0\rangle$,
corresponding to the contributions from Figs. 1 (a) and 1 (b),
respectively, and
\begin{eqnarray} I_{(^3P_0)}(\vec{P},m_1,m_2,m_3)&=&\int
d^3\vec{p}\,\mbox{}\psi^\ast_{n_BL_BM_{L_B}}
\left({\scriptstyle\frac{m_3}{m_1+m_3}}\vec{P}_B+\vec{p}\right)\psi^\ast_{n_CL_CM_{L_C}}
\left({\scriptstyle\frac{m_3}{m_2+m_3}}\vec{P}_B+\vec{p}\right)\nonumber\\
&&\times\psi_{n_AL_AM_{L_A}}
(\vec{P}_B+\vec{p}){\cal{Y}}^m_1(\vec{p}),
\end{eqnarray}
where $\vec{P}=\vec{P}_B=-\vec{P}_C$, $\vec{p}=\vec{p}_3$, $m_3$
is the mass of the created quark $q_3$, and the $\psi$'s are  the
relative wave functions in momentum space.

The spin overlap in terms of Wigner's $9j$ symbol can be given by
\begin{eqnarray}
&&\langle
\chi^{14}_{S_BM_{S_B}}\chi^{32}_{S_CM_{S_C}}|\chi^{12}_{S_AM_{S_A}}\chi^{34}_{1-m}\rangle=\nonumber\\
&&\sum_{S,M_S}\langle S_BM_{S_B}S_CM_{S_C}|SM_S\rangle\langle
S_AM_{S_A}1-m|SM_S\rangle\nonumber\\
&&\times(-1)^{S_C+1}\sqrt{3(2S_A+1)(2S_B+1)(2S_C+1)}\left\{\begin{array}{ccc}
\frac{1}{2}&\frac{1}{2}&S_A\\
\frac{1}{2}&\frac{1}{2}&1\\
S_B&S_C&S
\end{array}\right\}.
\end{eqnarray}

 In order to compare with the experiment conventionally,
${\cal{M}}^{M_{J_A}M_{J_B}M_{J_C}}_{(^3P_0)}(\vec{P})$ can be
converted into the partial amplitude by a recoupling
calculation\cite{recp}
\begin{eqnarray}
{\cal{M}}^{LS}_{(^3P_0)}(\vec{P})&=&
\sum_{\renewcommand{\arraystretch}{.5}\begin{array}[t]{l}
\scriptstyle M_{J_B},M_{J_C},\\\scriptstyle M_S,M_L
\end{array}}\renewcommand{\arraystretch}{1}\!\!
\langle LM_LSM_S|J_AM_{J_A}\rangle\langle
J_BM_{J_B}J_CM_{J_C}|SM_S\rangle\nonumber\\
&&\times\int
d\Omega\,\mbox{}Y^\ast_{LM_L}{\cal{M}}^{M_{J_A}M_{J_B}M_{J_C}}_{(^3P_0)}
(\vec{P}). \label{pwave}
\end{eqnarray}
If we consider the relativistic phase space, the decay width
$\Gamma_{(^3P_0)}(A\rightarrow BC)$ in terms of the partial wave
amplitudes is
\begin{eqnarray}
\Gamma_{(^3P_0)}(A\rightarrow BC)= \frac{\pi
P}{4M^2_A}\sum_{LS}|{\cal{M}}^{LS}_{(^3P_0)}|^2. \label{width1}
\end{eqnarray}
Here
$P=|\vec{P}|$=$\frac{\sqrt{[M^2_A-(M_B+M_C)^2][M^2_A-(M_B-M_C)^2]}}{2M_A}$,
and $M_A$, $M_B$, and $M_C$ are the masses of the meson $A$, $B$,
and $C$, respectively.

The simple harmonic oscillator (SHO) approximation for the meson
space wave functions is used. This is typical of decay
calculations and it has been demonstrated that using the more
realistic space wave functions, such as those obtained from
Coulomb, plus the linear potential model, does not change the
results significantly\cite{flux,3p0sho1,3p0sho2}. Under the SHO
wave function approximation, the partial amplitudes and partial
widths for $A\rightarrow BC$ can be calculated analytically based
on relations (\ref{pwave}) and (\ref{width1}), respectively.

In momentum space, the SHO wave function is
\begin{eqnarray}
\psi_{nLM_L}(\vec{p})=R^{\mbox{\tiny
SHO}}_{nL}(p)Y_{LM_L}(\Omega_p), \label{mspace}
\end{eqnarray}
where the radial wave function is given by
\begin{eqnarray}
R^{\mbox{\tiny SHO}}_{nL}(p)=\frac{(-1)^n(-i)^L}{\beta^{3/2}}
\sqrt{\frac{2n!}{\Gamma(n+L+\frac{3}{2})}}\left(\frac{p}{\beta}\right
)^L e^{(-p^2/2\beta^2)}L^{L+(1/2)}_n\left(
\frac{p^2}{\beta^2}\right).
\end{eqnarray}
Here $\beta$ is the SHO wave function scale parameter, and
$L^{L+(1/2)}_n\left(\frac{p^2}{\beta^2}\right)$ is an associated
Laguerre polynomial.

\subsubsection{The flux-tube model of meson decay} \indent
\vspace*{-1cm}

The flux-tube model is based on the strong-coupling Hamiltonian
lattice formulation of QCD\cite{flux}. In the flux-tube model, a
meson $A$ consists of a quark ($q_1$) and antiquark ($\bar{q}_2$)
connected by a tube of chromoelectric flux. Meson decay occurs
when the flux-tube breaks at a point along its length, and a
quark-antiquark pair ($q_3\bar{q}_4$) is created from the vacuum
to connect to the free ends of the flux-tubes, leaving a final
state consisting of two mesons $B$ and $C$.

The flux-tube model of meson decay is similar to the $^3P_0$
model, but extends the $^3P_0$ model by considering the actual
dynamics of the flux-tubes. This is done by including a flux-tube
overlap function that represents the overlap of the flux-tube of
the initial meson $A$ with those of the two outgoing mesons $B$
and $C$. The flux-tube overlap function reflects the spatial
dependence of the pair-creation amplitude. For the conventional
$q\bar{q}$ meson decay, the flux-tube overlap is usually chosen as
the following form\cite{flux}
\begin{eqnarray}
\gamma(\vec{r}_A,\vec{\mbox{y}})=\gamma_0
\exp\left(-\frac{1}{2}b{\vec{\mbox{y}}_{\perp}}^2\right).
\end{eqnarray}
Here $\gamma_0$ is the pair-creation constant, $b$ is the string
tension, $\vec{\mbox{y}}$ is the pair $(q_3\bar{q}_4)$ creation
position,
$\vec{\mbox{y}}_{\perp}=-(\vec{\mbox{y}}\times\hat{\vec{r}}_A)\times\hat{\vec{r}}_A$
, and $\vec{r}_A$ is the antiquark-quark axes of meson A (see Fig.
2).

 \begin{figure}[hbt]
\begin{center}
\epsfig{file=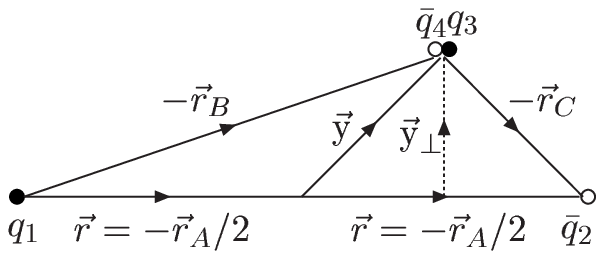,width=6.0cm, clip=}
\vspace*{0.5cm}\vspace*{-0.8cm}
 \caption{\small $A\rightarrow B+C$ in the flux-tube model.}
\end{center}
\end{figure}

The expression for
${\cal{M}}^{M_{J_A}M_{J_B}M_{J_C}}_{(\mbox{\tiny ft})}(\vec{P})$,
the amplitude of $A\rightarrow BC$ in the flux-tube model is the
same as that of relation (\ref{3p0exp}) except that the $\gamma$
is replaced by the $\gamma_0$ and
$I_{(^3p_0)}(\vec{P},m_1,m_2,m_3)$ is replaced by
\begin{eqnarray}
I_{(\mbox{\tiny
ft})}(\vec{P},m_1,m_2,m_3)&=&-\frac{8}{(2\pi)^{3/2}}\int
d^3\vec{r}\int
d^3\vec{\mbox{y}}\,\mbox{}\psi^\ast_{n_BL_BM_{L_B}}(\vec{r}_B)
\psi^\ast_{n_CL_CM_{L_C}}
(\vec{r}_C)\nonumber\\
&&\times{\cal{Y}}^m_1\left((\vec{P}+i\vec{\bigtriangledown}_{\vec{r}_A})\psi_{n_AL_AM_{L_A}}
(\vec{r}_A)\right
)\exp\left(-\frac{1}{2}b{\vec{\mbox{y}}_{\perp}}^2\right
)\nonumber\\
&&\times\exp\left(i\vec{P}\cdot(m_+\vec{r}+m_-\vec{\mbox{y}})\right
),
\end{eqnarray}
where $\vec{r}_A=-2\vec{r}$, $\vec{r}_B=-\vec{\mbox{y}}-\vec{r}$,
and $\vec{r}_C=\vec{\mbox{y}}-\vec{r}$ as shown in Fig. 2.
$m_+=\frac{m_1}{m_1+m_3}+\frac{m_2}{m_2+m_3}$,
$m_-=\frac{m_1}{m_1+m_3}-\frac{m_2}{m_2+m_3}$, and the $\psi$'s
are now the relative wave functions in position space.

As in the $^3P_0$ model, the SHO wave function approximation for
the meson space wave functions is taken. In position space, the
SHO wave function is the Fourier transform of (\ref{mspace})
\begin{eqnarray}
\psi_{nLM_L}(\vec{r})=R^{\mbox{\tiny
SHO}}_{nL}(r)Y_{LM_L}(\Omega_r), \label{pspace}
\end{eqnarray}
where the radial wave function is given by
\begin{eqnarray}
R^{\mbox{\tiny SHO}}_{nL}(r)=\beta^{\frac{3}{2}}
\sqrt{\frac{2n!}{\Gamma(n+L+\frac{3}{2})}}\left(\beta r\right )^L
e^{(-\beta^2 r^2/2)}L^{L+(1/2)}_n(\beta^2 r^2).
\end{eqnarray}

With these elements, the partial amplitudes and partial widths for
$A\rightarrow BC$ in the flux-tube model can also be calculated
analytically based on relations (\ref{pwave}) and (\ref{width1}),
respectively.

\subsection{ Decay properties of the $\eta_2(1870)$ as the
$\eta_2(2Dn\bar{n})$}
\indent \vspace*{-1cm}

Under the SHO wave function approximation, the parameters used in
this work involve the SHO wave function scale parameter $\beta$,
the pair production strength parameter $\gamma$ in the $^3P_0$
model, the pair-creation constant $\gamma_0$ and the string
tension $b$ in the flux-tube model, and the constituent quark mass
$m_q$. In this work, we choose to follow the
Refs.\cite{3p0rev4,3p0sho2,prd72} and take $\gamma=8.77$,
$\beta_A=\beta_B=\beta_C=\beta=0.4$ GeV, $\gamma_0=14.3$, $b=0.18$
GeV$^2$, $m_u=m_d=0.33$ GeV, and $m_s=0.55$ GeV which are also the
values used to evaluate the decays of the
$\pi_2(1880)$\cite{lidm2009}\footnote{Our value of $\gamma$ is
higher than that used by Ref.\cite{prd72} (0.505) by a factor of
$\sqrt{96\pi}$, due to different field conventions, constant
factor in $T$, etc. The calculated results of the widths are, of
course, unaffected.}. The meson masses used to determine the phase
space and final state momenta are\footnote{ We assume that the
$a_0(1450)$, $f_0(1370)$, and $K^\ast_0(1430)$ are the ground
scalar meson as Refs.\cite{strangebarnes,nnhybridbarnes}.}
$M_{\pi}=138$ MeV, $M_{\eta}=548$ MeV, $M_K=496$ MeV,
$M_{\rho}=776$ MeV, $M_{\omega}=783$ MeV, $M_{\phi}=1019$ MeV,
$M_{K^\ast}=894$ MeV, $M_{\pi(1300)}=1300$ MeV,
$M_{K^\ast(1410)}=1414$ MeV, $M_{a_1(1260)}=1230$ MeV,
$M_{f_1(1285)}=1282$ MeV, $M_{b_1(1235)}=1230$ MeV,
$M_{h_1(1170)}=1170$ MeV, $M_{K_1(1270)}=1272$ MeV,
$M_{K_1(1400)}=1403$ MeV, $M_{a_2(1320)}=1318$ MeV,
$M_{f_2(1270)}=1275$ MeV, $M_{K^\ast_2(1430)}=1429$ MeV,
$M_{a_0(1450)}=1474$ MeV, $M_{f_0(1370)}=1370$ MeV, and
$M_{K^\ast_0(1430)}=1425$ MeV. The meson flavor wave functions
follow the conventions of Ref.\cite{strangebarnes}.  Based on
(\ref{width1}), the numerical values of the partial decay widths
of the $\eta_2(1870)$ as the $\eta_2(2Dn\bar{n})$ are listed in
Table 1.

{\small
\begin{table}[hbt]
\begin{center}
\vspace*{-0.5cm}
 \caption{\small Partial widths of the $\eta_2(1870)$ as the
$\eta_2(2Dn\bar{n})$ in the $^3P_0$ model and the flux-tube model(
in MeV). The initial state mass is set to $1842$ MeV.}
 \vspace*{0.5cm}

\begin{tabular}{l|l|l}\hline
Mode  & $\Gamma_{LS}$ in $^3P_0$ model & $\Gamma_{LS}$  in
flux-tube model
\\\hline

$K^\ast K$   & $\Gamma_{P1}=13.95$  &  $\Gamma_{P1}=15.19$ \\
            & $\Gamma_{F1}=3.75$  &$\Gamma_{F1}=4.09$  \\\hline
$\rho\rho$  &  $\Gamma_{P1}=43.00$  & $\Gamma_{P1}=46.83$\\
             & $\Gamma_{F1}=9.16$  &$\Gamma_{F1}=9.97$   \\\hline
$\omega\omega$&  $\Gamma_{P1}=14.23$  &$\Gamma_{P1}=15.50$\\
              & $\Gamma_{F1}=2.66$  &  $\Gamma_{F1}=2.90$\\\hline
$K^\ast K^\ast$ & $\Gamma_{P1}=2.08$     & $\Gamma_{P1}=2.27$   \\
                & $\Gamma_{F1}=0.01$  & $\Gamma_{F1}=0.01$ \\\hline
$ K_1(1270)K$ &$\Gamma_{D1}=0.06$ &$\Gamma_{D1}=0.07$\\\hline

$a_1(1260)\pi$& $\Gamma_{D1}=15.23$&$\Gamma_{D1}=16.59$\\\hline
$f_1(1285)\eta$ &$\Gamma_{D1}=0.00$ &$\Gamma_{D1}=0.00$\\\hline

$a_2(1320)\pi$ &$\Gamma_{S2}=67.56$ &$\Gamma_{S2}=73.58$\\
               & $\Gamma_{D2}=34.52$  & $\Gamma_{D2}=37.59$  \\
               &$\Gamma_{G2}=0.43$&$\Gamma_{G2}=0.47$\\\hline
$ f_2(1270)\eta$ &$\Gamma_{S2}=17.47$ &$\Gamma_{S2}=19.02$\\
                 &$\Gamma_{D2}=0.02$&$\Gamma_{D2}=0.02$\\
                 &$\Gamma_{G2}=0.00$&$\Gamma_{G2}=0.00$\\\hline
$a_0(1450)\pi$ &$\Gamma_{D0}=2.37$ &$\Gamma_{D0}=2.58$\\\hline
$\Gamma$     & 226.50       &246.68\\\hline
\end{tabular}
\end{center}
\end{table}
}

 It is clear from Table
1 that the numerical results in the $^3P_0$ model are similar to
those in the flux-tube model. The total width of the
$\eta_2(2Dn\bar{n})$ at 1842 MeV is expected to be about 226 MeV
in the $^3P_0$ model or about 246 MeV in the flux-tube model, both
compatible with the $\eta_2(1870)$ width. The expected dominant
decay modes are $a_2(1320)\pi$, $\rho\rho$, $f_2(1270)\eta$,
$a_1(1260)\pi$, $\omega\omega$, and $K^\ast K$, in accord with the
$\eta_2(1870)$ dominantly decaying to $a_2(1320)\pi$ and
$f_2(1270)\eta$. It should be noted that the partial width of
$\eta_2(2D n\bar{n})\rightarrow K K^\ast$ is expected to be large
($\Gamma(K K^\ast)/\Gamma(f_2(1270)\eta)\approx 1$), however,
there is no indication of the $\eta_2(1870)$ in the WA102 data for
the $KK\pi$\cite{wwaa}, which suggests that the $\eta_2(1870)$
does not decay significantly to $K K^\ast$. This discrepancy could
arise from the omission of the small $n\bar{n}\leftrightarrow
s\bar{s}$ flavor mixing effect in the $\eta_2(1870)$. With
$\eta_2(1870)\equiv \cos\theta n\bar{n}-\sin\theta s\bar{s}$,
where $\theta$ is the mixing angle, in the $^3P_0$ model, the
dependence of the predicted total width $\Gamma(\eta_2(1870))$ and
the partial widths for the dominant decay modes on the mixing
angle $\theta$ are shown in Fig. 3. (The results from the
flux-tube model are very similar to those from the $^3P_0$ model).
Fig. 3 indicates that the $\Gamma(K K^\ast)$ is very sensitive to
the $\theta$. For small and negative $\theta$
($\theta\simeq-0.3\sim -0.2$ radians), the total width and other
dominant partial widths of the $\eta_2(1870)$ shown in Table 1 are
not significantly changed but the $\Gamma(KK^\ast)$ would be
small.
 \vspace*{1.5cm}
 \begin{figure}[hbt]
 \begin{center}
\vspace{-1.5cm}
 \epsfig{file=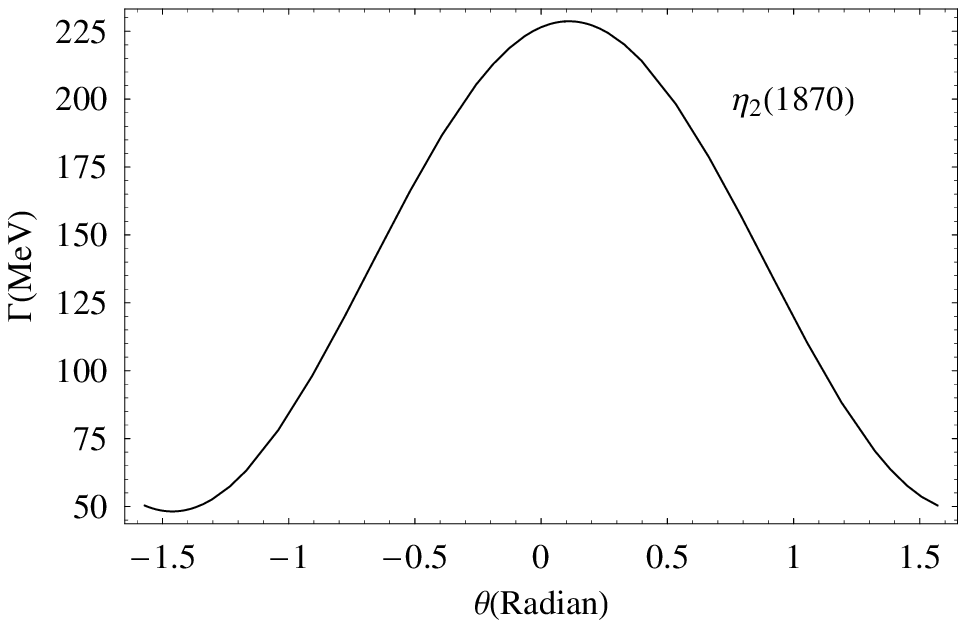,width=5.0cm, clip=}
\epsfig{file=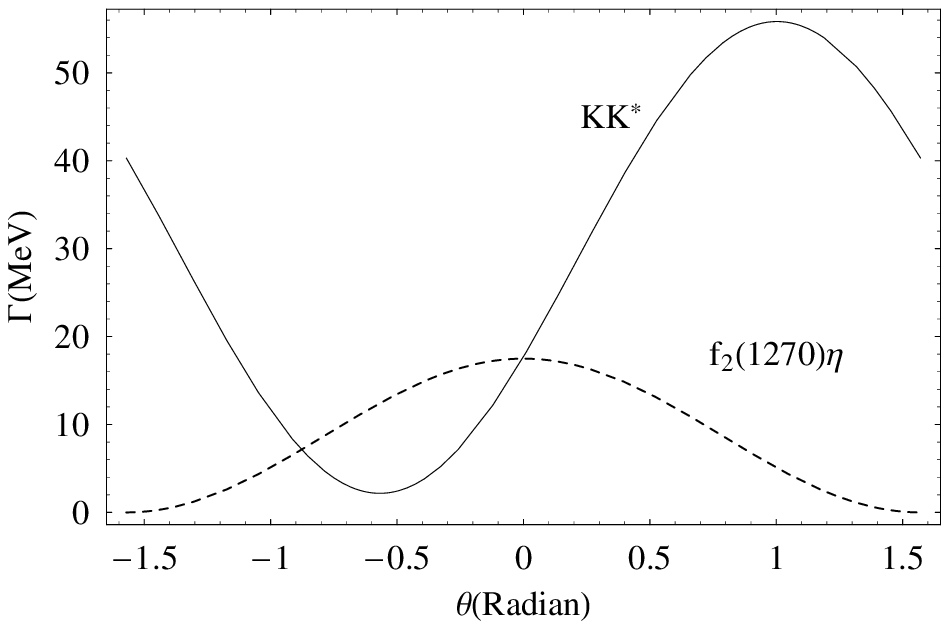,width=4.9cm, clip=}
\epsfig{file=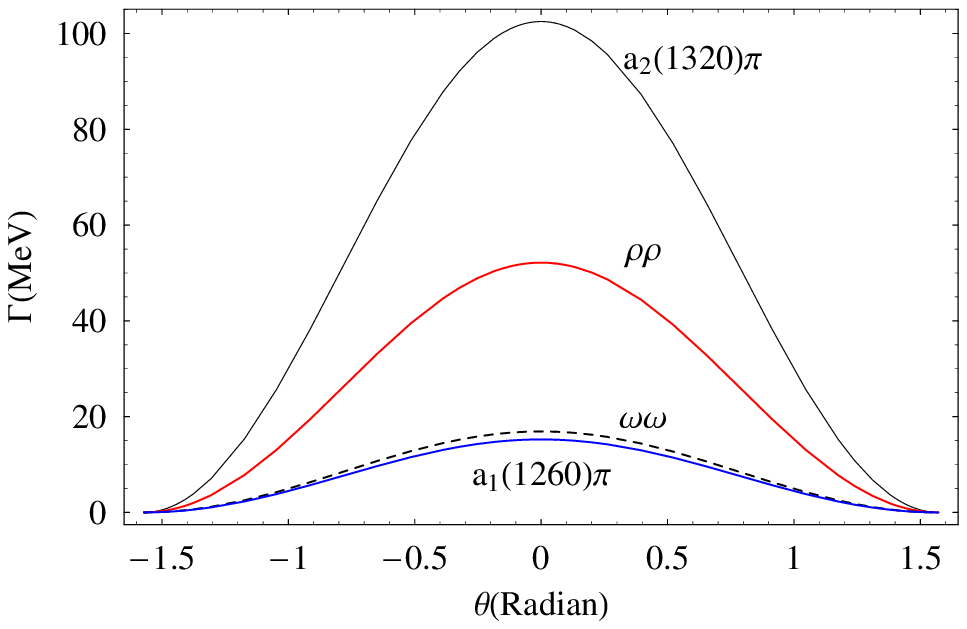,width=5.0cm, clip=}
 \vspace*{0.5cm}\vspace*{-1cm}
 \caption{\small In the $^3P_0$ model, the predicted  $\Gamma(\eta_2(1870))$, $\Gamma(a_2(1320)\pi)$, $\Gamma(\rho\rho)$, $\Gamma(\omega\omega)$, $\Gamma(a_1(1260)\pi)$, $\Gamma(K K^\ast)$ and $\Gamma(f_2(1270)\eta)$ versus the mixing angle
 $\theta$.
 }
\end{center}
\end{figure}

The decay dynamics of the $\eta_2(H n\bar{n})$ with a mass of
1.8-2.0 GeV has been investigated in the flux-tube
model\cite{close1,nnhybridbarnes,pss}. We now shall compare the
hybrid and quarkonium assignments for the $\eta_2(1870)$. Both
assignments lead to the significant $a_2(1320)\pi$ and
$f_2(1270)\eta$ signals, in accord with the experiment. The most
characteristic decay modes are the $\rho\rho$ and $\omega\omega$,
which are forbidden for the $\eta_2(H n\bar{n})$ due to the
selection rule while significant for the $\eta_2(2D n\bar{n})$.
Similar result follows for the $a_1(1260)\pi$. The $\rho\rho$ and
$\omega\omega$ channels would be the strong discriminant between
the hybrid and conventional meson for the $\eta_2(1870)$.
Unfortunately, the experimental information on the $\rho\rho$ and
$\omega\omega$ channels for the $\eta_2(1870)$ is not available.
Also, the value of $R=\Gamma(a_2(1320)\pi)/\Gamma(f_2(1270)\eta)$
for the $\eta_2(2D n\bar{n})$ is in fact different from that for
the $\eta_2(H n\bar{n})$. For example, at 1875 MeV, we predicted
$R=4$ for the $\eta_2(2D n\bar{n})$ while Barnes et al. predicted
$R=8$ for the $\eta_2(H n\bar{n})$\cite{nnhybridbarnes}.
Experimentally, the Crystal Barrel Collaboration gave $R=4.1\pm
2.3$\cite{ppb}, the WA102 Collaboration gave $R=20.4\pm
6.6$\cite{wa1022}, and Anisovich et al. gave $R=1.27\pm
0.17$\cite{ppb1}. The world average value of $R$ quoted by PDG is
$6\pm 5$\cite{pdg08}\footnote{The PDG does not quote the datum of
$1.27\pm 0.17$, and does not use this datum for averages, limits,
etc. }. Obviously, the uncertainty of the world average value for
the $R$ is so large that we can not distinguish the $\eta_2(2D
n\bar{n})$ assignment from the $\eta_2(H n\bar{n})$ interpretation
for the $\eta_2(1870)$ based this ratio. The further confirmation
of this ratio is needed. At present, the total width and the
strong decay pattern of the $\eta_2(1870)$ do not exclude the
possibility of it being in fact the $2\,^1D_2$ $q\bar{q}$ state.

\section{Discussions}
\indent\vspace{-1cm}

As mentioned in Sect.1, in $\bar{p}p$ annihilation the
$I^G(J^{PC})=0^{+}(2^{-+})$ resonance called $\eta_2(2030)$ has
been observed in the $a_2(1320)\pi$ and $f_2(1270)\eta$
channels\cite{ppb1,ppb2}. This state is listed as `Further state'
by PDG\cite{pdg08}. To some extent, the discovery of the
$\eta_2(2030)$ leads to the conjecture of the $\eta_2(1870)$ being
a hybrid because the $\eta_2(2030)$ looks like the first radial
excitation of the $\eta_2(1645)$ based on its mass. We shall turn
to the possibility of the $\eta_2(2030)$ being the $\eta_2(2D
n\bar{n})$ by investigating its decay dynamics. The partial widths
of the $\eta_2(2030)$ as the $\eta_2(2D n\bar{n})$ are estimated
in the $^3P_0$ model and flux-tube model. The numerical results
are listed in Table 2. The predictions from the $^3P_0$ model are
similar to those from the flux-tube model. We find if the
$\eta_2(2030)$ is the $\eta_2(2D n\bar{n})$, its total width would
be about 654 MeV or 709 MeV, far more than the experiment.
Therefore, the $\eta_2(2D n\bar{n})$ assignment for the
$\eta_2(2030)$ seems unfavorable in the $^3P_0$ model and the
flux-tube model. We also estimate the partial widths of the
$\eta_2(2030)$ as the $I=0$, $3\,^1D_2$ $n\bar{n}$ state
[$\eta_2(3D n\bar{n})$] in Table 2.  If the $\eta_2(2030)$ is the
$\eta_2(3D n\bar{n})$, we find (1) its total width would be about
150 MeV or 162 MeV; (2) it dominantly decays to $a_2(1320)\pi$
$D$-wave; (3) $B(f_2\eta)/B(a_2\pi)_{L=2}$ is about 0.1. All these
predictions are consistent with
experiment\cite{ppb1,ppb2}\footnote{The one exception to this is
that our predicted $B(a_2\pi)_{L=0}/B(a_2\pi)_{L=2}$ for the
$\eta_2(2030)$ as the $\eta_2(3D n\bar{n})$ is about 0.01,
inconsistent with the experiment of about $0.74\pm
0.17$\cite{ppb1}.}, and therefore the $\eta_2(2030)$ is more
likely to be the candidate for the $\eta_2(3D n\bar{n})$. This
indicates that the presence of the $\eta_2(2030)$ does not
contradict with the $\eta_2(2D n\bar{n})$ interpretation for the
$\eta_2(1870)$.

We note that the $\eta_2(1645)$, $\eta_2(1870)$, and
$\eta_2(2030)$\footnote{The masses of these three states are 1617
MeV, 1842 MeV, and 2030 MeV, respectively\cite{pdg08}.}
approximately populate a common trajectory as shown in Fig. 4. The
quasi-linear trajectories at the $(n, M^2)$-plots turned out to be
able to described the light mesons with a good
accuracy\cite{regge}.  Fig. 4 therefore indicates that the
$\eta_2(1870)$ and $\eta_2(2030)$ could be in fact the $2\,^1D_2$
and $3\,^1D_2$ $q\bar{q}$ states and the narrow level spacing
between the $\eta_2(2D n\bar{n})$ and $\eta_2(3D n\bar{n})$ is not
surprising in Regge phenomenology. More recently, Li and
Chao\cite{chao1,chao2} have found that the coupled-channel and
screening effects are important for the spectra of higher
charmonia, and the masses of higher charmonia from the screened
potential model or coupled-channel model are considerably lower
than those from the naive quark model. The coupled-channel effect
or the screening effect may be also a factor leading to the narrow
level spacing between the $\eta_2(2D n\bar{n})$ and $\eta_2(3D
n\bar{n})$. More theoretical investigations and more complete data
on the light mesons are needed to clarify this issue.

\vspace{1.5cm}
 \begin{figure}[hbt]
 \begin{center}
\vspace{-1.5cm}
 \epsfig{file=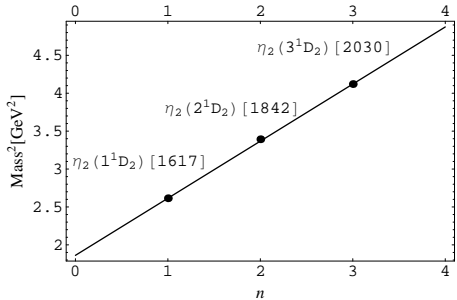,width=6.0cm, clip=}
 \vspace*{0.5cm}\vspace*{-1cm}
 \caption{\small The $(n,M^2)$-plot for the $\eta_2(1645)$, $\eta_2(1870)$, and
 $\eta_2(2030)$.
 }
\end{center}
\end{figure}

With the $\eta_2(2030)$ being the $\eta_2(3D n\bar{n})$, one can
expect that the isovector member of the $3\,^1D_2$ meson nonet
would lie around 2030 MeV, which leads to that the $\pi_2(2005)$
with a mass of $2005\pm 15$ MeV and a width of $200\pm 40$
MeV\cite{pi2005} could be a good candidate for the $I=1$,
$3\,^1D_2$ $q\bar{q}$. We find if the $\pi_2(2005)$ is the
$3\,^1D_2$ $q\bar{q}$, its total width would be about 122 MeV in
the $^3P_0$ model or 130 MeV in the flux-tube model, roughly
consistent with the experiment.

{\small
\begin{table}[hbt]
\begin{center}
\vspace*{-0.5cm}
 \caption{\small Partial widths of the $\eta_2(2030)$ as the $\eta_2(2D n\bar{n})$ and $\eta_2(3D n\bar{n})$
 in the $^3P_0$ model and the flux-tube model (in MeV). The initial state mass is set to $2030$ MeV.}
 \vspace*{0.5cm}

\begin{tabular}{l|ll|ll}\hline
      &\multicolumn{2}{c}{$\eta_2(2D n\bar{n})$}&\multicolumn{2}{c}{$\eta_2(3D
      n\bar{n})$}\\\hline
Mode  &$^3P_0$ model& flux-tube model& $^3P_0$ model& flux-tube
model\\\hline

$K^\ast K$&27.80       & 30.27   &5.00  &5.44\\
$\rho\rho$&80.83       &88.02    &13.14 &14.31\\
$\omega\omega$&26.42   &28.77    &4.28  &4.66\\
$K^\ast K^\ast$&14.22  &15.49    &2.88  &3.13\\
$K^\ast(1410)K$&6.32   &3.92    &5.38  &4.46\\
$K_1(1270)K$&1.68      &1.83    &0.36  &0.39\\
$K_1(1400)K$&0.59      &0.64    &0.14  &0.15\\
$b_1(1235)\rho$&152.98 &166.60    &35.91 &39.11\\
$h_1(1170)\omega$&65.10&70.89    &13.60  &14.81\\
$a_0(1450)\pi$&16.82   &18.32    &9.26  &10.09\\
$f_0(1370)\eta$&0.60   &0.65    &0.38  &0.41\\
$K^\ast_0(1430)K$& 0.59&0.64    &0.32  &0.34\\
$a_1(1260)\pi$&41.11   &44.77    &5.71  &6.22\\
$f_1(1285)\eta$& 1.43  &1.56    &0.29  &0.31\\
$a_2(1320)\pi$&129.60   &141.14    &35.00  &38.11\\
$f_2(1270)\eta$&23.47  &25.56    &4.21  &4.58\\
$K^\ast_2(1430)K$&64.78     &70.54    &14.23  &15.49\\
$\Gamma$       & 654.34&709.61    &150.09  &162.01\\\hline
&\multicolumn{4}{c}{Experiment: $\Gamma_{\eta_2(2030)}$=$205\pm
10\pm 25$\cite{ppb1} or $190\pm 40$\cite{ppb2}}\\\hline
\end{tabular}
\end{center}
\end{table}
}

Generally speaking, the pure $\eta_2(2D n\bar{n})$ can mix with
the pure $\eta_2(H n\bar{n})$ to produce the physical state. We
shall discuss the possibility of the $\eta_2(1870)$ being a
mixture of the $\eta_2(2D n\bar{n})$ and $\eta_2(H n\bar{n})$.
 Obviously, quantitative determination of its
$q\bar{q}$-hybrid content should be essential to confirm or refute
this possibility. The available decay information for the
$\eta_2(1870)$ is unfortunately not sufficient to do
this\footnote{Within the $\eta_2(1870)$ being the mixture of the
$\eta_2(2D n\bar{n})$ and $\eta_2(H n\bar{n})$, the measured
partial widths of the $\eta_2(1870)$ are needed to determine its
hybrid-quarkonium content quantitatively.}. However, we can
qualitatively estimate the hybrid component of the $\eta_2(1870)$
would be small if the $\eta_2(1870)$ is really a mixture of the
$q\bar{q}$ and hybrid.  As mentioned in Sect. 3, the fact of the
$\eta_2(1645)$ and $\eta_2(1870)$ having the same behavior as a
function of the $dp_T$\cite{wa1021,wa1022} strongly suggests the
$\eta_2(1870)$ having the same dynamical structure as the
$\eta_2(1645)$, which makes the substantial hybrid admixture in
the $\eta_2(1870)$ unlikely. The further experimental information
of the $\eta_2(1870)$ in the $\rho\rho$ and $\omega\omega$
channels would be crucial to shed light on this issue.

\section{Summary and conclusion}
\indent \vspace*{-1cm}

From the mass, production, total width, and strong decay pattern
of the $\eta_2(1870)$, we point out that the possibility of it
being a canonical $2\,^1D_2$ $q\bar{q}$ state does exist. Also,
the decay information for the $\eta_2(2030)$ is consistent with it
being a $3\,^1D_2$ rather than $2\,^1D_2$ $q\bar{q}$ state, and
the total width of the $\pi_2(2005)$ favors the argument that it
could be the candidate for the isovector partner of the
$\eta_2(2030)$. The possibility of the $\eta_2(1870)$ being a
mixture of hybrid and $q\bar{q}$ might exist while the substantial
hybrid admixture in this state seems unlikely. The further
experimental information of the $\eta_2(1870)$ in the $\rho\rho$
and $\omega\omega$ channels is needed. We tend to conclude that
the $\eta_2(1870)$ is the ordinary $2\,^1D_2$ $q\bar{q}$ state or
the $2\,^1D_2$ $q\bar{q}$ with small hybrid admixture, as does the
$\pi_2(1880)$.

 \section*{Acknowledgments}
We thank K. T. Chao for useful suggestions and discussions. We
also thank D. V. Bugg for helpful comments. This work is supported
in part by HANCET under Contract No. 2006HANCET-02, and by the
Program for Youthful Teachers in University of Henan Province.

\end{document}